\documentstyle[11pt,ihepconf,twoside,epsfig]{article}

\def\be{\begin{equation}}
\def\ee{\end{equation}}
\def\mbf#1{\mbox{\boldmath $#1$}}

\input{amssym.def}
\input{amssym}

\input{amssym.def}
\input{amssym}

\begin{document}

  \title{ \Large\bf  QUANTIZATION of PHYSICAL MODELS \\[1mm] and 
NON-COMMUTATIVE GEOMETRY}
 
 \author{{\bf P.A. Saponov}\\
 {\small\it Institute for High Energy Physics, 142200, Protvino, Russia}}
 
 \date{}
  \maketitle

\section{Introduction}

\par 
In my talk I would like to give a simple introduction
into some ideas of the non-commutative (NC) geometry. At
present the NC geometry is a fairly wide and quickly
developing branch of mathematics which has close connections
to physics.

Loosely speaking, the main subject of the NC geometry is a
"manifold" whose "coordinates" cannot be presented by the
ordinary numbers but belong to a non-commutative algebra.
Actually, such like objects are quite usual for physics ---
they appear in the quantization of physical systems.

Indeed, let us recall the standard scheme of canonical
quantization (leaving apart the numerous subtleties and
problems connected with some concrete systems). We begin
with a real manifold $M$ called a configuration space. In
simple cases its (local) coordinates $\{q_a\}$ represent the
position of our system in the space. Then we pass to the
cotangent bundle $T^*M$ --- a phase space of the system, the
fibre coordinates $\{p_a\}$ being the corresponding momenta.
We are interested in the algebra of smooth real functions
${\cal A} = C^\infty (T^*M)$ on the cotangent bundle which is
a commutative algebra of dynamical variables. The time
evolution is defined by a Hamiltonian and a Poisson structure
on $T^*M$. Due to the Darboux theorem, the Poisson structure
an be locally brought to the canonical form
\be
\{q_a, p_b\} = \delta_{ab}, \quad \{q_a,q_b\} =
\{p_a,p_b\} = 0.
\label{can-pb}
\ee

The canonical quantization consists in passing from the
commutative algebra $\cal A$ to some non-commutative algebra
${\cal A}_{\hbar}$ generated by the elements $\hat q_a$ and
$\hat p_a$ with the commutation relations
\be
[\hat q_a, \hat p_b] = i\hbar\,\delta_{ab}, \quad [\hat q_a,
\hat q_b] = [\hat p_a,\hat p_b] = 0.
\label{can-q}
\ee

From the mathematical point of view, this procedure can be 
interpreted as a transition from the commutative geometry to
the NC one. Indeed, a theorem of algebraic geometry \cite{Sh}
tells that an affine variety $M$ is completely defined by the
algebra $\cal A$ of regular (polynomial) functions on $M$ (the
coordinates of a point is a particular example of such
functions). Therefore, the non-commutative algebra
${\cal A}_\hbar$ (being sought of as an algebra of functions)
defines some non-commutative ``manifold". The quotation marks
stand here in order to stress that the non-commutative (or
quantum) ``manifold" cannot be presented as a set of points in a
space~--- a visual image is lost here.

The canonical quantization scheme described above needs a
significant modification if the phase space cannot be
covered by a single coordinate chart. In this case the canonical
coordinates (\ref{can-pb}) cannot be fixed globally on the phase
space. The typical and important example of such a situation
is a {\it constrained system}, that is a system whose motion
is somehow restricted. A simple example is given by a particle
moving on the surface of a sphere.

In the most cases, the phase space of a constrained system is a
submanifold $M\subset T^*V$, $V$ being an $n$-dimensional linear
space, defined by the set of relations
\be
\Phi_s(q,p) = 0,\quad s=1,2,\dots, r,
\label{sur-cons}
\ee
which are called {\it the constraints}. The main idea of the
Dirac quantization scheme for the constrained systems consists
in a modification of the Poisson structure on $T^*V$ in such a
way that it would be equivalent to its restriction on the
surface of constraints (\ref{sur-cons}). It allows us to work
with the initial $(q,p)$ coordinates and set the constraints
equal to zero in the strong sense \cite{Di}. When quantizing the
system, one should impose the commutation relations on $\hat q$
and $\hat p$ in accordance with these modified (or Dirac)
brackets.

However, the Dirac brackets are often nonlinear in coordinates
and the problems of ordering appears. The problem of the
ordering intertwines with the fact that the quantum algebra of
observables ${\cal A}_\hbar$ is an {\it associative} algebra
(not a Lie algebra). Therefore, we must define the {\it product}
$\hat F_1\cdot \hat F_2$ (not only the commutator) of any two
elements $\hat F_1$ and $\hat F_2$. But the canonical
commutation relations fix only an {\it antisymmetrical} part of
the product and we have to use some additional physical reasons
in order to restore the whole quantum algebra and choose the
correct ordering in the brackets. In practice the product of
quantum operators for the constrained system can be defined only
as a (formal) series in the Planck constant $\hbar$.

In my talk I consider the quantization of the algebra of
functions on a semisimple orbit of fhe coadjoint action of the
general linear group $GL(n)$. For any connected Lie group $G$
such an orbit is a $G$-homogeneous symplectic submanifold
\cite{Ki} in the linear space $g^*$ dual to the Lie algebra $g$.
Being symplectic manifold, the orbit can be treated as a phase
space of a physical system with the symmetry group $G$.
Moreover, any homogeneous symplectic manifold, whose symmetry
group is a connected Lie group $G$, is locally isomorphic to
some orbit of the coadjoint action of $G$ (or the coadjoint
action of its central extension $G_1$ \cite{Ki}). This fact
shows the importance of the orbits and their universality
in the problem of quantization of the physical systems whose
symmetries form a Lie group $G$.

\section{The Poisson brackets and orbits of a Lie group}

In this section we recall some notions of the Hamiltonian
formalism and the theory of coadjoint orbits of a Lie group
(for detail see \cite{Ar, Ki}).

\subsection{The Hamiltonian mechanics}

Consider a real manifold M and let
${\cal A} = C^{\infty}(M,\Bbb R)$ be the set of smooth real
functions on $M$. With respect to the pointwise multiplication,
addition and multiplication by numbers the set ${\cal A}$ forms
a commutative algebra. The elements of $\cal A$ are called {\it
the dynamical variables}.

Suppose, that the manifold $M$ is endowed with a Poisson
structure. This means that there exists a bilinear operation
$\{\;,\}:{\cal A}\times {\cal A}\rightarrow {\cal A}$, called
a {\it Poisson bracket}, which satisfies the following
requirements:
\begin{itemize}
\item $\{f_1,f_2\} = -\{f_2,f_1\}$;
\item $\{f_1,f_2f_3\} = \{f_1,f_2\}f_3 + \{f_1,f_3\}f_2$;
\item $\{f_1,\{f_2,f_3\}\} +{\rm cycle}(1,2,3) = 0$.
\end{itemize}

The time evolution of a dynamical variable $f$ is defined by
a Hamiltonian $H\in{\cal A}$ in accordance with the formula
$$
\dot f = \{f,H\}.
$$

Denote by $z_a$ the set of (local) coordinates on $M$ and
let
\be
\{z_a,z_b\} = \omega_{ab}(z).
\label{om}
\ee
The requirements on the Poisson brackets listed above transform
into the following properties of the tensor $\omega$
$$
\begin{array}{rll}
{\bf i)} & \omega_{ab}(z) = -\omega_{ba}(z)\quad&
\mbox{the skew-symmetry} \ ,\\
\rule{0pt}{5mm}
{\bf ii)} & \omega_{as}(z)\partial^s\omega_{bc}(z) +
{\rm cycle}(a,b,c) = 0\quad&\mbox{the Jacobi identity} \ .
\end{array}
$$
The Poisson bracket of two functions is now written in the
explicit form
$$
\{f,g\}(z) = \frac{\partial f}{\partial z_a}\,\omega_{ab}(z)
\,\frac{\partial g}{\partial z_b}.
$$

The Poisson bracket is called {\it non-degenerate} if
\be
\{f,g\} \equiv 0\quad \forall\,g\in{\cal A}
\quad \Leftrightarrow\quad  f\equiv 0.
\label{nd-pb}
\ee
This is equivalent to the invertibility of $\omega_{ab}$
$$
\exists \,\omega^{ab}(z):\quad
\omega^{ac}(z)\omega_{cb}(z) = \delta^a_b.
$$
In this case $M$ is a symplectic manifold. The corresponding
closed non-degenerate two form is given by
$$
\Omega = \omega^{ab}(z)\,dz_a\wedge dz_b.
$$
This is a common situation in mechanics --- the manifold $M$ is
often chosen to be the cotangent bundle of some configuration
space $V$ and therefore is a symplectic manifold.

\subsection{The Poisson-Lie brackets}

The simplest case of (\ref{om}) corresponds to the constant
tensor $\omega$. We shall consider the next step in
complexity when the manifold $M$ is a finite dimensional vector
space and the tensor $\omega$ is a linear function in
coordinates
\be
\omega_{ab}(z) = C_{ab}^sz_s.
\label{lin-om}
\ee
Taking into account the properties {\bf i)} and {\bf ii)}
of $\omega$ we come to the following restrictions on the
coefficients $C_{ab}^c$
\begin{eqnarray}
&& C_{ab}^c = - C_{ba}^c\label{a-C}\\
&&\rule{0pt}{5mm}
\sum_sC_{ab}^sC_{sc}^r +{\rm cycle}(a,b,c) = 0.
\label{J-C}
\end{eqnarray}
As is well known, relations (\ref{a-C}) and (\ref{J-C})
means that the coefficients $C_{ab}^c$ form the set of
structure constants of some Lie algebra $g$. The space $M$
can be identified with its dual space $g^*$. A Poisson
bracket with the tensor $\omega$ given by (\ref{lin-om})
is called the {\it Poisson-Lie bracket}.

In what follows we shall consider the case $g=gl(n,\Bbb R)$. Let
us recall that the Lie algebra $gl(n,\Bbb R)$ is generated by
the $n^2$ elements $e_{ij}$, $1\le i,j\le n$, subject to the
following relations
\be
[e_{ij},e_{rs}] = \delta_{jr}\,e_{is} - \delta_{is}e_{rj}.
\label{gln}
\ee
The elements $e_{ij}$ form a basis of the algebra and an
arbitrary element $\mbf{a}\in g$ is in one-to-one correspondence
with the $n\times n$ matrix of its coefficients 
$$
\mbf{a}\in g\quad\leftrightarrow\quad
A = \|a_{ij}\|
\in{\rm Mat}_n(\Bbb R):\quad
\mbf{a} = \sum_{i,j}a_{ij}e_{ij}.
$$

The dual space $g^*$ is by definition a space of linear
functionals on $g$. If $g$ is finite dimensional, then $g^*$ can
be described as a linear vector space endowed with a
non-degenerated bilinear form
${\cal h}\;,\;{\cal i}:\,g^*\times g\rightarrow \Bbb R$.
In our example we shall fix in $g^*$ the {\it dual} basis
$\epsilon_{ij}$ defined by the following pairings
$$
{\cal h}\epsilon_{ij},e_{rs}{\cal i} = \delta_{jr}\delta_{is}.
$$
An arbitrary linear functional on $g$ is also represented
by the matrix of its coefficients with respect to the basis
$\epsilon_{ij}$
$$
\mbf{x}\in g^*\quad\leftrightarrow\quad
X = \|x_{ij}\|\in {\rm Mat}_n(\Bbb R):\quad
\mbf{x} = \sum_{i,j}x_{ij}\epsilon_{ij}.
$$
Now the value of a linear functional $\mbf{x}\in g^*$ on an
arbitrary element $\mbf{a}\in g$ can be written in the form
\be
{\cal h}\mbf{x},\mbf{a}{\cal i} = {\rm Tr}(XA)\quad
\mbf{x}\in g^*,\;\mbf{a}\in g.
\label{dual-p}
\ee
The smooth functions on $g^*$ are those in the coordinates
$x_{ij}$
$$
f(\mbf{x}) = f(x_{ij}),\quad \mbf{x} =
\sum_{i,j}x_{ij}\epsilon_{ij}.
$$
The linear space $g^*$ is a Poisson manifold, the Poisson-Lie
bracket is defined by the structure constants of (\ref{gln})
\be
\{x_{ij}, x_{rs}\} = \delta_{jr}x_{is} - \delta_{is}x_{rj}.
\label{gln-pb}
\ee
The subset of {\it linear} functions on $g^*$ forms a Lie algebra
with respect to the above bracket and this algebra is isomorphic
to the initial Lie algebra $gl(n,\Bbb R)$.

The Poisson bracket (\ref{gln-pb}) is degenerate. Consider the
set of functions $p_{k}$ defined as follows
\be
p_{k}(\mbf{x}) = {\rm Tr}(X^k), \quad X = \|x_{ij}\|.
\label{def:pk}
\ee
Since  $\{x_{ij}, p_k(x)\} = 0$, the functions $p_k$ has zero
Poisson bracket with any function $f\in C^\infty(g^*,\Bbb R)$.
In other words, the functions $p_k$ are central elements of the
infinite dimensional Lie algebra $C^{\infty}(g^*,\Bbb R)$ with
respect to the Poisson-Lie bracket (\ref{gln-pb}). In
accordance with definition (\ref{nd-pb}) this means that
(\ref{gln-pb}) is degenerate.

\subsection{The orbits of the coadjoint representation}

For a Lie group $G$ there exists an important representation
in the linear space of its Lie algebra $g$ which is called the
{\it adjoint representation}. In our example the Lie group
$GL(n,{\Bbb R})$, corresponding to $g=gl(n,\Bbb R)$, can be
identified with the group of invertible real $n\times n$
matrices. By definition the adjoint representation reads
$$
{\rm Ad}_{M}(\mbf{a})=
\sum_{ij}(MAM^{-1})_{ij}e_{ij},\quad M\in G,\quad
\mbf{a} = \sum_{ij}a_{ij}e_{ij}\in g, \;A=\|a_{ij}\|.
$$
With the help of a non-degenerate bilinear form ${\cal h}
\,,{\cal i}$ one can define the {\it coadjoint} representation
${\rm Ad}^*_M$ of $G$ in the dual space $g^*$. By definition
$$
{\cal h} {\rm Ad}^*_M(\mbf{x}),\mbf{a}{\cal i} =
{\cal h}\mbf{x},{\rm Ad}_{M^{-1}}(\mbf{a}){\cal i}
$$
Using (\ref{dual-p}) we can find the coordinate matrix $X'$
of the element ${\rm Ad}^*_M(\mbf{x})$
$$
{\rm Tr}(X'A) = {\rm Tr}(XM^{-1}AM) = {\rm Tr}(MXM^{-1}A).
$$
Since $A$ is an arbitrary matrix, we come to the following
form of the {\it coadjoint} action of $G$ on $g^*$
\be
{\rm Ad}^*_M(\mbf{x}) = \mbf{x}'\quad\Rightarrow\quad
X' = MXM^{-1}.
\label{co-act}
\ee

From now on, we shall identify the space $g^* = gl(n,\Bbb R)^*$
with the space ${\rm Mat}_n(\Bbb R)$ of the $n\times n$ real
matrices.

Let us fix a matrix $X\in g^*$ and consider the subset
${\cal O}_X\subset g^*$ formed by the following matrices
\be
{\cal O}_X = \Big\{MXM^{-1}\;|\;M\in G\Big\}.
\label{def:orb}
\ee
The set ${\cal O}_X$ is called the {\it orbit} of the
coadjoint representation (action) of $G$ passing through $X$.

Let us list some properties of the orbits.

\noindent
{\bf a)}
Any orbit ${\cal O}_X$ is a $G$-homogeneous set. This means
that $G$ acts transitively on ${\cal O}_X$: any two
points $X_1,X_2\in {\cal O}_X$ can be mapped into each
other by some element of $G$. Indeed, if $X_1 = M_1XM_1^{-1}$,
$X_2 = M_2XM_2^{-1}$ then
$$
X_1 = (M_1M_2^{-1})X_2 (M_1M_2^{-1})^{-1}.
$$
Therefore, any point of an orbit can be taken as its
representative element.
\medskip

\noindent
{\bf b)}
Any orbit is a symplectic $G$-invariant manifold \cite{Ki}.
This theorem comprises several important statements.

b1) An orbit ${\cal O}_X$ is not simply a set, it is a
submanifold in $g^*$.
Below we explain how one can define an orbit by a set of
polynomial equations in coordinates of $g^*$.

b2) Being restricted to the algebra of functions on ${\cal O}_X$,
the Poisson bracket (\ref{gln-pb}) is non-degenerated.

b3) The action of $G$ on ${\cal O}_X$ is Poisson. This means
the following. The coadjoint action of $G$ on the space $g^*$
defined by (\ref{co-act}) is extended to the functions $f$ on
$g^*$ in the standard way
$$
f\stackrel{M}{\longrightarrow} f_M:\quad
f_M(\mbf{x}) = f({\rm Ad}^*_{M^{-1}}(\mbf{x}))\quad M\in G,\;
\mbf{x}\in g^*.
$$
This action is Poisson if it commutes with the calculation
of the Poisson bracket
$$
\{f_1,f_2\}_M = \{{f_1}_M, {f_2}_M\}.
$$
\noindent
{\bf c)} An orbit is invariant under the diffeomorphisms
generated by an {\it arbitrary} strictly Hamiltonian vector
field on $g^*$. Therefore, if the initial state of a dynamical
system is represented by a point on an orbit $\cal O$, then
during the time evolution with {\it any} Hamiltonian
$H\in C^\infty(g^*,\Bbb R)$ the states of this system are
constrained to the same orbit $\cal O$.

This is a quite general result valid for any connected Lie
group (see, for example, \cite{Ar,Ol}). Below we give a
simple proof of this statement for the (semisimple) orbits
of $GL(n,\Bbb R)$. For this purpose, we should analize
relations which define an orbit. This is a subject of the
next subsections.

\subsection{Generic semisimple orbits}

We shall restrict ourselves to the class of {\it semisimple}
orbits. The orbit is called semisimple if it contains a matrix,
which can be diagonalized by a similarity transformation (a
semisimple matrix by definition). In accordance with the
property {\bf a)}, all the points of a semisimple orbit
are semisimple matrices.

Fix a diagonal matrix $X = {\rm diag}(\mu_1,\dots,\mu_n)\in g^*$
and consider its orbit ${\cal O}_X$ under the coadjoint action
(\ref{co-act}) of the group $G$. Let us first suppose that
all eigenvalues $\mu_i$ are pairwise distinct numbers. Such an
orbit will be called {\it generic}. As is known from linear
algebra, any $n\times n$ matrix $X$ satisfies the polynomial
Cayley-Hamilton identity of the form
\be
\sum_{k=0}^n (-X)^{n-k}\sigma_k(X)\equiv 0, \quad X^0 = I,
\label{CH-cl}
\ee
$I$ being the $n\times n$ unit matrix. The coefficient
$\sigma_k(X)$ is the sum of all principal minors of $X$ of the
$k$-th order. In terms of the spectral values $\mu_i$ of $X$ the
coefficient $\sigma_k(X)$ is the $k$-th order elementary
symmetric function in $\mu_i$
$$
\sigma_k(X) = \sum_{1\le i_1<\dots <i_k\le n}
\mu_{i_1}\dots \mu_{i_k},\quad 1\le k \le n.
$$
The functions $p_k(X)$ defined in (\ref{def:pk}) (recall, that
we identify $\mbf{x}$ with the corresponding matrix $X$) are the
power sums of the eigenvalues of $X$
$$
p_k(X) = \sum_{i=1}^n\mu_i^k.
$$
The functions $\sigma_k$ and $p_k$ are connected by the well
known Newton relations
$$
k\sigma_k -p_1\sigma_{k-1}+p_2\sigma_{k-2} -\dots
+(-1)^kp_k \equiv 0, \quad 1\le k\le n.
$$
With these relations one can easily prove that
$\{x_{ij},\sigma_k(X)\}\equiv 0$ as soon as we have
$\{x_{ij},p_k(X)\}\equiv 0$.

The elements $p_k(X)$, $1\le k\le n$ are algebraically
independent, as well as $\sigma_k(X)$, $1\le k\le n$. The
higher power sums $p_r(X)$ at $r>n$ are polynomials in the
first $n$ functions $p_k$. This follows from~(\ref{CH-cl})
and Newton relations.

Let us now find a set of relations, which define the orbit
${\cal O}_X$ as a submanifold in $g^*$. For this purpose,
observe, that the coadjoint action (\ref{co-act}) preserve
the spectrum of $X$. Due to this fact, all the points of
${\cal O}_X$ have the same eigenvalues $\mu_i$ and, therefore,
the same values of the central functions
$p_k(X)$, $1\le k\le n$. So, the necessary condition for a
matrix $Y\in g^*$ to belong to the orbit ${\cal O}_X$ is
$$
Y\in {\cal O}_X \;\Rightarrow\; p_k(Y)=p_k(X),\quad
k=1,2,\dots,n
$$
Since we consider a semisimple generic orbit (all eigenvalues
are pairwise distinct) this condition is also sufficient.

Therefore, a semisimple generic orbit in $gl(n,\Bbb R)^*$ is
defined by the set of $n$ polynomial relations
\be
p_k(X) = {\rm Tr}(X^k) = c_k,\quad c_k = \sum_{i=1}^n\mu_i^k,
\quad 1\le k\le n,
\label{gen-orb}
\ee
where the numbers $\mu_i$, $1\le i\le n$, are pairwise distinct.

At last, since $p_k$ are central functions with respect to the
Poisson-Lie bracket (\ref{gln-pb}), we get
$$
\{p_k(X),H(X)\} \equiv 0, \quad \forall H\in
C^{\infty}(g^*,\Bbb R).
$$
This means that a Hamiltonian vector field $\xi_H = \{\;,H\}$,
generated by an arbitrary Hamiltonian $H$, is tangent to any
generic semisimple orbit, defined by (\ref{gen-orb}). This
proves the property {\bf c)} for this type of orbits.

\subsection{Generic orbits in $gl(2,\Bbb R)^*$}

Let us consider a simple example of generic orbits in
$gl(2,\Bbb R)^*$. This is a space of $2\times 2$ real valued
matrices which is isomorphic (as a vector space) to
${\Bbb R}^4$. The Cayley-Hamilton identity reads
$$
X^2 - X{\rm Tr}(X) +I\det X\equiv 0, \quad
X=\left(\!\!
\begin{array}{cc}
x_{11} & x_{12}\\
x_{21} & x_{22}
\end{array}\!\!
\right).
$$
Fix two different number $\mu_1$ and $\mu_2$. In order they
could represent the spectrum of a matrix $X\in gl(2,\Bbb R)^*$,
the numbers $\mu_1$ and $\mu_2$ must either be both real or
be complex conjugate to each other: $\bar \mu_1 = \mu_2$.

Then a generic orbit is a submanifold in ${\Bbb R}^4$ defined
by a couple of relations
\be
x_{11} + x_{22} = \mu_1+\mu_2,\quad
x_{11}x_{22} - x_{12}x_{21} = \mu_1\mu_2.
\label{ex:gen-2}
\ee
Parametrizing $x_{ij}$ by the new coordinates $x$, $y$ and $t$
$$
x_{11} = \frac{\mu_1+\mu_2}{2}+x,\quad
x_{22} = \frac{\mu_1+\mu_2}{2}-x,\quad
x_{12} = y-t,\quad
x_{21} = y+t
$$
we reduce system (\ref{ex:gen-2}) to the following equation
\be
t^2 - x^2 - y^2 = - \frac{1}{4}\,(\mu_1-\mu_2)^2.
\label{eq:orb-2}
\ee
If $\mu_1$ and $\mu_2$ are real, this equation defines
a one sheet hyperboloid around the $t$-axis.

If the eigenvalues are complex conjugate then the above
equation transforms to
$$
t^2 - x^2 - y^2 = ({\rm Im}\mu_1)^2.
$$
This is an equation of a two sheet hyperboloid around the
$t$-axis.

Note, that a cone, corresponding to the zero right hand side
in (\ref{eq:orb-2}), is not a generic orbit (since
$\mu_1=\mu_2$).

\subsection{The non-generic orbits}

Consider now the case, when the representative matrix $X$
of the orbit ${\cal O}_X$ possesses coincident eigenvalues.
Suppose that the spectrum of $X$ consists of $r<n$ pairwise
distinct numbers $\mu_i$, each of them entering the spectrum
with a multiplicity $m_i\ge 1$
$$
{\rm Spec}(X) = \{(\mu_i,m_i)_{\,1\le i\le r}\,|
\,m_1+m_2+\dots +m_r = n\}.
$$
The functions $p_k$ take the values
\be
p_k(X)-c_k =0,\quad c_k = \sum_{i=1}^rm_i\mu_i^k.
\label{pk-mult}
\ee
But now their fixation does not define a semisimple orbit.
Indeed, if we choose the values of $p_k(X)$ as in
(\ref{pk-mult}), the Cayley-Hamilton identity (\ref{CH-cl})
can be written in the factorized form
\be
(X-\mu_1I)^{m_1}(X-\mu_2I)^{m_2}\dots
(X-\mu_rI)^{m_r}\equiv 0.
\label{CH-mult}
\ee
If all multiplicities $m_i =1$ (the generic case) then identity
(\ref{CH-mult}) guarantees the matrix $X$ to be semisimple.
Therefore, on fixing $p_k$ as in (\ref{gen-orb}), one uniquely
defines (up to a permutation of eigenvalues) a semisimple matrix
$X$ with a spectrum $\mu_i$.

If there are the multiplicities $m_i>1$ then (\ref{CH-mult})
has several solutions with the same spectrum, but only one
of them is a semisimple matrix.

Consider an example for the case $n=3$. Let the spectrum
consists of two eigenvalues \mbox{$(\mu_1, m_1 =1)$} and
$(\mu_2, m_2 =2)$. The Cayley-Hamilton identity (\ref{CH-mult})
looks as follows
\be
(X-\mu_1I)(X-\mu_2I)^2\equiv 0
\label{CH-3}
\ee
and it has two non-equivalent solutions
$$
X_1 = \left(\!
\begin{array}{ccc}
\mu_1 & 0 & 0 \\
0 & \mu_2 & 0 \\
0 & 0 & \mu_2
\end{array}
\!\right)
\quad{\rm and}\quad
X_2 = \left(\!
\begin{array}{ccc}
\mu_1 & 0 & 0 \\
0 & \mu_2 & 1 \\
0 & 0 & \mu_2
\end{array}
\!\right).
$$
Evidently, only the matrix $X_1$ defines a semisimple orbit, but
$p_k(X_1) = p_k(X_2)$ for all $k$.

So, relations (\ref{pk-mult}) where some $m_i>1$ define a
{\it union} of orbits and only one orbit of them is semisimple.
That is why we should find some additional relations in order
to uniquely extract a semisimple orbit.

This problem is easy to solve. Let us pay attention to the fact
that the semisimple matrix $X_1$ in the above example actually
obeys a polynomial identity of a {\it lower} order than that of
identity (\ref{CH-3}). Indeed
\be
(X_1 - \mu_1I)(X_1-\mu_2I) \equiv 0.
\label{CH-2}
\ee
This is the so called {\it minimal polynomial identity} for
the matrix $X_1$. As is known from matrix analysis, a matrix $X$
is semisimple if and only if its minimal polynomial is a product
of linear factors with pairwise distinct spectral values and
with all exponents to be equal to unity:
\be
\mbox{$X$ is semisimple}\quad \Leftrightarrow \quad
(X-\mu_1I)\dots (X-\mu_rI)\equiv 0.
\label{min-pol}
\ee
So, in order to extract a semisimple non-generic orbit we
should complete (\ref{pk-mult}) with $n^2$ new relations
\be
\Phi_{ij}(X) = 0,\quad 1\le i,j\le n,
\label{phi-rel}
\ee
where $\Phi_{ij}$ are the matrix elements of (\ref{min-pol})
$$
\Phi_{ij}(X) = (X^r)_{ij} - \sum_{a=1}^r\mu_a\,(X^{r-1})_{ij}
+\dots + (-1)^r\,\delta_{ij}\,\mu_1\dots\mu_r.
$$

Since the minimal polynomial identity (\ref{min-pol}) is of
the $r$-th order, then in (\ref{pk-mult}) only first $r$
functions $p_k$ should be fixed (others are polynomials
in these $p_k$-s).

It is worth stressing, that in general we cannot disregard
relations\footnote{It is only possible in the most degenerate
case when the spectrum consists of the single eigenvalue $\mu$.
Then the multiplicity is uniquely defined by the size of the
matrix and the only semisimple matrix is proportional to the
unit matrix $X=\mu I$.} (\ref{pk-mult}). Having fixed the
identity~(\ref{min-pol}) alone, we again define a union of
orbits, each of them being semisimple. The representative
matrices of these orbits have the same spectral values but
different multiplicities. For example, the identity (\ref{CH-2})
is satisfied by the two non-equivalent semisimple matrices
$$
X_1 = \left(\!
\begin{array}{ccc}
\mu_1 & 0 & 0 \\
0 & \mu_2 & 0 \\
0 & 0 & \mu_2
\end{array}
\!\right)
\quad{\rm and}\quad
X_1' = \left(\!
\begin{array}{ccc}
\mu_1 & 0 & 0 \\
0 & \mu_1 & 0 \\
0 & 0 & \mu_2
\end{array}
\!\right).
$$
The role of relations (\ref{pk-mult}) consists in fixing
multiplicities $m_i$ of spectral values $\mu_i$.

The question on the invariance of the non-generic orbit under
the Hamiltonian evolution is slightly more involved in comparing
with the generic case. The matter is that the matrix elements
$\Phi_{ij}$ of the identity (\ref{min-pol}) do not
Poisson-commute with  the coordinates $x_{ij}$ (in contrast with
the power sums $p_k$). Nevertheless, the {\it full} set of
relations (\ref{phi-rel}) defines a manifold which is invariant
under the diffeomorphisms of the Hamiltonian vector fields. To
prove this we first calculate the bracket
$\{x_{ks}, (X^m)_{ij}\}$ on the base of (\ref{gln-pb})
$$
\{x_{ks}, (X^m)_{ij}\}  = \delta_{si}(X^m)_{kj} - \delta_{kj}
(X^m)_{is},\quad X=\|x_{ij}\|.
$$
Then, taking into account the structure of $\Phi_{ij}$, we come
to the analogous result for the bracket of $x_{ks}$ and
$\Phi_{ij}$
\be
\{x_{ks}, \Phi_{ij}\}  = \delta_{si}\Phi_{kj} - \delta_{kj}
\Phi_{is}.
\label{F-inv}
\ee
This result is quite obvious except for the question about the
constant term proportional to the product $\mu_1\dots \mu_r$.
Being constant, the term has vanishing Poisson bracket with
$x_{ks}$ and at the first glance it is not clear, how it can
appear in the right hand side of (\ref{F-inv}). The mater is
that the constant term is only contained in the diagonal matrix
elements $\Phi_{ii}$. A simple analysis shows that the right
hand side of (\ref{F-inv}) depends on the {\it difference} of
diagonal matrix elements. The only bracket with such a
dependence is of the form
$$
\{x_{ji}, \Phi_{ij}\}  = \Phi_{jj} - \Phi_{ii}.
$$
The constant terms are cancelled and the right hand side of
(\ref{F-inv}) is actually independent of them.

Now it is a straightforward consequence of (\ref{F-inv}) that the
manifold defined by the system of relations (\ref{phi-rel}) is
tangent to a Hamiltonian vector field generated by an arbitrary
Hamiltonian $H\in C^{\infty}(g^*,\Bbb R)$.

\subsection{Physical systems on orbits of Lie groups}

There are known a plenty of examples of mechanical
systems the phase space of which can be presented as an
orbit of some Lie group. A great number of them can be found,
for example, in \cite{Pe, FT} with the detailed consideration.

Here we turn to a few of such examples just for the illustration
of the above consideration.
\medskip

\noindent
{\bf Example 1.} Take the Lie group $G=SO(3)$ --- the group of
rotations of the three dimensional Euclidean space ${\Bbb R}^3$.
The Poisson-Lie structure on the space $g^*=so(3)^*$ is
$$
\{x_i, x_j\} = \epsilon_{ijk}x_k,
$$
where $\epsilon_{ijk}$ are the components of the full
antisymmetric tensor.

The orbits of the coadjoint $SO(3)$-action are the two
dimensional spheres ${\cal O}_r$
$$
{\cal O}_r:\quad x_1^2+x_2^2+x_3^2 = r^2.
$$

Given a Hamiltonian $H(x)$, the dynamical equations read
$$
\frac{d\vec x}{dt} = {\vec x}\times {\vec\nabla}H.
$$
The quadratic Hamiltonian $H=\frac{1}{2}\sum_{i}J_ix_i^2$ leads
to the Euler equation describing the rotation of a rigid body
with the principle moments of inertia $J_i$ around a fixed
point. The vector $\vec x$ represents the angular velocity of
the body.

\medskip

\noindent
{\bf Example 2 \cite{Pe}.} The Lie group $G=E(3)$ is a group
of motions of the three dimensional Euclidean space
${\Bbb R}^3$. This group is a semidirect product of the group
of rotations and the group of translations.

The dual space to its Lie algebra $g^* = {\rm Lie}^*(E(3))$
is a six dimensional vector space with coordinates
$(x_i, y_i)$ obeying to the following bracket relations
$$
\{x_i,x_j\} = \epsilon_{ijk}x_k, \quad
\{x_i,y_j\} = \epsilon_{ijk}y_k, \quad
\{y_i,y_j\} = 0.
$$

The orbits are four dimensional manifolds parametrized
by the two numbers $a\ge 0$ and $b$
$$
{\cal O}_{a,b} = \{\vec x, \vec y \,:\quad\vec x\cdot \vec y
 = ab,\;|\vec y|^2 = a^2,\; a\ge 0\}.
$$
A quadratic Hamiltonian leads to the Kirchhoff equations for
the motion of a rigid body in an ideal liquid.

\section{Quantization of algebra of functions on an orbit}

In this section we extend the field of real numbers to the
complex field $\Bbb C$. This means that we shall work with
complex manifolds and shall consider complex valued functions
on the manifolds. Besides, we constrain ourselves to the
{\it regular} functions on a manifold. By definition, a
function is called regular on a manifold $M$ if it coincides
with a polynomial in each coordinate chart of the manifold. The
set of all regular functions on a complex manifold $M$ is called
a {\it coordinate ring} of $M$ and is denoted as ${\Bbb C}[M]$.
Of course, it is also an algebra over $\Bbb C$.

Let us consider a problem of quantization of the algebra of
functions over an orbit ${\cal O}_X$ of the coadjoint action
of the Lie group $G=GL(n)$ on the space $g^* = gl(n,\Bbb C)^*$.
The construction of the orbit presented in the previous section
can be extended to the complex case in a straightforward way.

Consider the main features of the quantization from the physical
point of view. Let we are given a system $S$ and let $\cal A$ be
the algebra of its dynamical variables (observables) endowed
with a Poisson structure. The algebra $\cal A$ is an associative
commutative algebra with respect to pointwise multiplication and
addition of functions and is a (infinite dimensional) Lie
algebra with respect to the Poisson bracket.

To quantize a system $S$ means the following:

\begin{enumerate}

\item One should pass from the commutative associative algebra
$\cal A$ to some non-commutative associative algebra
${\cal A}_\hbar$ which is called a quantized algebra of
dynamical variables (observables). In practice one usually works
not with the algebra ${\cal A}_\hbar$ itself but with some its
representation in a Hilbert space, the observables being
represented by hermitian operators.

\item
Being a Lie algebra with respect to the commutator, the
quantum algebra should be isomorphic to the Lie algebra
obtained from $\cal A$ with the help of the Poisson structure.
That is, if $f\rightarrow \hat f$ and $g\rightarrow \hat g$
then
$$
\{f,g\}\rightarrow \frac{1}{i\hbar}\,[\hat f,\hat g].
$$

\item 
A quasiclassical limit
$$
\lim_{\hbar\rightarrow 0} {\cal A}_\hbar =\cal A
$$
should be defined (in some sense).

\item 
At the quantization the number of degrees of freedom must not
changed and the symmetries of the classical system should be
maximally retained.
\end{enumerate}
The second point is central in the above scheme and causes the
main difficulties. The matter is that even if the brackets of
generators have the simplest form (\ref{can-pb}) then there is
an ambiguity in the construction of the correspondence
$f\rightarrow \hat f$ -- the well known problem of ordering.
In general case we can represent a quantum "function"
$\hat f$ as a (formal) series in the Planck constant
$\hat f = \sum \hbar^s\hat f_s$ and try to find the coefficients
$\hat f_s$ in such a way that the correspondence described
in the point 2 could be satisfied. This means, that generally the
associative multiplication in the quantum algebra
${\cal A}_\hbar$ can be restored only up to a finite order in
$\hbar$.

The mathematical definition of the quantization procedure
accepted in the non-commutative geometry is very close to the
physical one.

Given a coordinate ring ${\cal A} = \Bbb C[M]$ of some affine
algebraic variety\footnote{This simply means that a variety is
defined by a system of polynomial equations in some affine
space.} $M$ we should pass to the non-commutative algebra
${\cal A}_\hbar$ parametrized by a formal parameter $\hbar$
in such a way that the following requirements should be
satisfied:
\begin{itemize}
\item
The initial algebra $\cal A$ should be isomorphic to the
following quotient 
$$
{\cal A}_\hbar/\hbar{\cal A}_\hbar\cong {\cal A}
$$
and the first order term of the quantum product must be defined
by the Poisson bracket in $\cal A$
$$
\hat a\star\hat b = ab +\frac{\hbar}{2}\{a,b\} +
o(\hbar^2).
$$
This is analogous to the points 2 and 3 of the physical scheme.
\item
The quantization should be a {\it flat} deformation.
This means that the "supply" of elements of ${\cal A}_\hbar$
is as large as that of $\cal A$. This condition is equivalent
to the physical requirement that the quantization should not
alter the number of degrees of freedom of a system.
\item
If the algebra $\cal A$ is a module over some another algebra
(or a group) $\cal B$ then the quantum algebra ${\cal A}_\hbar$
should also be a module over $\cal B$ (or, possibly, over
some its deformation\footnote{The quantum groups is one of
the most known examples of such a deformation.}). This is
equivalent to the physical requirement that the symmetries of
the classical system (represented by $\cal B$) should be
extended to the quantum case.
\end{itemize}

Let us shortly describe the result of quantization of the
algebra of regular functions on a semisimple orbit ${\cal O}_X$
of the coadjoint action of the Lie group $G=GL(n)$. The
importance of the result consists, in particular, in the fact
that the quantum algebra can be described explicitly as a
quotient of the universal enveloping algebra $U(gl(n))$ over
some ideal generated by a polynomial relations (not as a series
in the quantization parameter $\hbar$).

Let us fix a matrix $X\in gl(n)^*$ with the spectrum
\be
{\rm Spec}(X) = (\mu_i,m_i),\quad 1\le i\le r.
\label{X-spec}
\ee
Recall, that the eigenvalues $\mu_i$ are all distinct and
$m_i\ge 1$ are the corresponding multiplicities. Consider then
the orbit ${\cal O}_X$ of the coadjoint action of $G$. The
number $r\le n$ is called the rank of the orbit. As was
explained in Section 2, the orbit is defined by the set of
relations (\ref{pk-mult}) and (\ref{phi-rel}).

Turn now to the coordinate ring ${\Bbb C}[g^*]$ of $g^*$ which
is a $\Bbb C$-algebra of all polynomials in the coordinates
$x_{ij}$. Let us construct a two-sided ideal ${\cal J}_X$ in
${\Bbb C}[g^*]$ generated by the elements $\Phi_{ij}$ and
$\pi_k = p_k-c_k$. This ideal is a set of all elements of
${\Bbb C}[g^*]$ which can be presented in the form
$$
{\cal J}_X = \{a\Phi_{ij}b, \;g\pi_kh\;|\;a,b,g,h\in
{\Bbb C}[g^*], \;1\le i,j\le n,\;1\le k\le r\}.
$$
Then as is well known (see, for example, a textbook \cite{Sh}),
the algebra of regular functions on the orbit ${\cal O}_X$ is
given by the following quotient
$$
{\Bbb C}[{\cal O}_X] = {\Bbb C}[g^*]/{\cal J}_X.
$$

After the quantization of the Poisson brackets (\ref{gln-pb})
the algebra ${\Bbb C}[g^*]$ turns into the universal enveloping
algebra $U(gl(n)_\hbar)$. The coordinates $x_{ij}$ map into the
$U(gl(n)_\hbar)$ generators $e_{ij}$ with the commutation
relations
$$
[e_{ij},e_{ks}] = \hbar(\delta_{jk}e_{is} - \delta_{si}e_{kj}).
$$

Consider the matrix $L=\|e_{ij}\|$ composed of the
non-commutative generators $e_{ij}$. It can be shown, that the
matrix $L$ satisfies the Cayley-Hamilton identity analogous to
(\ref{CH-cl}) but with modified coefficients \cite{GS}. The
quantities $p_k = {\rm Tr}(L^k)$ are central
elements of $U(gl(n)_\hbar)$.

Let us first suppose that all the multiplicities are equal to
unity (the generic orbit). Define a {\it quantum spectrum} as
$n$ numbers $\nu_i = \mu_i$ (that is in the generic case the
quantum spectrum coincides with the classical one). Then
consider the two-sided ideal ${\cal J}(\nu)$ in 
$U(gl(n))_\hbar$ which is generated by the set of $n$ elements
$$
\pi_k = {\rm Tr}(L^k) - c_k, \quad
c_k = \sum_{i=1}^n\nu_i^kd_i,
$$
where the {\it quantum multiplicities} are not equal to unity
even in the generic case
$$
d_i = \prod_{j=1\atop j\not=i}^n\
\frac{(\nu_i - \nu_j - \hbar)}{\nu_i-\nu_j}.
$$
Then the quantized algebra ${\cal A}_\hbar $ of functions
on the generic orbit ${\cal O}_X$ turns out to be the
following quotient of the $U(gl(n)_\hbar)$ (see \cite{GS2})
$$
{\cal A}_\hbar = U(gl(n)_\hbar)/{\cal J}(\nu).
$$

Consider now the non-generic case when there exist
multiplicities greater than 1. With each pair $(\mu_i,m_i)$
we associate a {\it string} of {\it quantum eigenvalues}
$$
(\mu_i,m_i)\;\rightarrow\;(\mu_i,\,\mu_i + \hbar,\,\mu_i+
2\hbar,\, \dots,\;\mu_i+(m_i-1)\hbar).
$$
Note that all the elements of the string are distinct.

Then we define the quantum spectrum as the set of all elements
of all strings
$$
\nu_{i,s} = \mu_i + s\hbar, \quad 1\le i\le r,\;0\le s\le
m_i-1.
$$
That is the degenerate classical eigenvalues split into the set
of non-degenerate quantum spectral values.

The quantized algebra ${\cal A}_\hbar$ of functions on a
semisimple orbit ${\cal O}_X$ passing through the matrix
(\ref{X-spec}) is the quotient of $U(gl(n)_\hbar)$ over the
two sided ideal (see \cite{DM,GS2}) generated by the matrix
elements of the polynomial identity
$$
(L-\mu_1I)\dots (L-\mu_rI) = 0
$$
and by the elements
$$
\pi_k = {\rm Tr}(L^k)-c_k,\quad
c_k = \sum_{i,s}\nu_{i,s}^kd_{i,s},\quad 
1\le k\le r.
$$
The quantum multiplicities
$$
d_{i,s} = \prod_{(j,r)\not=(i,s)}\
\frac{(\nu_{i,s} - \nu_{j,r} - \hbar)}{\nu_{i,s}-\nu_{j,r}}
$$
are correctly defined since the quantum spectrum is not 
degenerated. 

It is worth noting, that only the multiplicities $d_{i,0}$ are
non-zero. All the numbers $d_{i,s}$ with $s\ge 1$ vanish due to
the structure of the quantum spectral values $\nu_{i,s}$. This
means that only the first elements $\mu_i$ of each string
contribute to the value of the trace 
$$
{\rm Tr}(L^k) = \sum_{i=1}^r\mu_i^kd_{i,0} \ .
$$

\section*{Conclusion}

In my talk I considered the simplest linear brackets
$\{\,,\}_{PL}$ defined by (\ref{gln-pb}). But on any semisimple
orbit in $gl(n)^*$ there exists another Poisson structure
$\{\,,\}_r$ connected with the classical R-matrix
$r\in g\wedge g$ \cite{DGK}. This Poisson structure is
compatible with (\ref{gln-pb}) and allows us to construct the so
called Poisson pencil on the orbit, that is a family of brackets
$$
\{\,,\}_{a,b} = a\{\,,\}_{PL} + b\{\,,\}_r
$$
with some constants $a$ and $b$. The quantization of such
a Poisson pencil leads to an orbit in the reflection equation
algebra connected with the corresponding quantum R-matrix $R$.

The reflection equation algebra plays a significant role in
the theory of integrable systems with boundaries and in the
non-commutative geometry. The detailed consideration of
the orbits in this algebra is given in \cite{GS2}.\\

\end{document}